\documentclass[aps,nofootinbib]{revtex4}
\usepackage{amsfonts,color,bbm,amsmath,amssymb}
\usepackage{graphicx,calc,epsfig}

\definecolor{blue}{rgb}{0,0,1}
\definecolor{green}{rgb}{0,1,0}
\definecolor{red}{rgb}{1,0,0}
\definecolor{van}{rgb}{1,0,1}
\definecolor{al}{rgb}{1,1,0}

\newcounter{mnotecount}[section]

\newcommand{\be}{\nopagebreak[3]\begin{equation}}
\newcommand{\ee}{\end{equation}}
\newcommand{\ba}{\nopagebreak[3]\begin{eqnarray}}
\newcommand{\ea}{\end{eqnarray}}
\DeclareFontFamily{U}{rsfs}{}         
\DeclareFontShape{U}{rsfs}{m}{n}{<5> rsfs5 <6><7> rsfs7          %
  <8><9><10><10.95><12><14.4><17.28><20.74><24.88> rsfs10}{}     %
\DeclareMathAlphabet{\mathfs}{U}{rsfs}{m}{n}                     %
\newcommand{\mfs}[1]{\mathfs {#1}}                               %

\def\cc{{\cal C}}

\def\cg{{\cal G}}

\def\cp{{C}}

\def\cs{{\cal S}}

\def\cv{{\cal V}}

\def\nnR{\nonumber\\}

\def\wg{\wedge}
\def\hgst{{}^*\!}

\newcommand{\va}{\scriptscriptstyle}

\newcommand{\sD}{{\mfs D}}

\newcommand{\sF}{{\mfs F}}

\newcommand{\sL}{{\mfs L}}

\newcommand{\inter}{{\lrcorner}}

\usepackage{psfrag}
\usepackage{vmargin}
\usepackage{amsfonts}
\usepackage{amsmath}
\usepackage{graphics,graphicx,color,calc,epsfig}
\usepackage{euscript}

\newcommand{\R}{\mathbb{R}}

\begin{document}

\title{A topological limit of gravity admitting an $SU(2)$ connection formulation}

\author{Lihui Liu$^1$}
\email{dliulhd@gmail.com}

\author{Merced Montesinos$^{1,2}$}
\email{merced@fis.cinvestav.mx}

\author{Alejandro Perez$^1$}
\email{perez@cpt.univ-mrs.fr}

\affiliation{$^1$Centre de Physique Th\'eorique\footnote{Unit\'e Mixte
de Recherche (UMR 6207) du CNRS et des Universit\'es Aix-Marseille
I, Aix-Marseille II, et du Sud Toulon-Var; laboratoire afili\'e
\`a la FRUMAM (FR 2291)}, Campus de Luminy, 13288 Marseille,
France.}

\affiliation{$^2$Departamento de F\'{\i}sica, Cinvestav, Instituto
Polit\'ecnico Nacional 2508, San Pedro Zacatenco, 07360, Gustavo A.
Madero, Ciudad de M\'exico, M\'exico.}

\date{\today \vbox{\vskip 2em}}

\begin{abstract}

We study the Hamiltonian formulation of the generally covariant theory
defined by the Lagrangian 4-form $\sL=e_I \wedge e_J \wedge F^{IJ}(\omega)$
where $e^I$ is a tetrad field and $F^{IJ}$ is the curvature of a
Lorentz connection $\omega^{IJ}$. This theory can be thought of as the
limit of the Holst action for gravity for the Newton constant $G\to \infty$ and Immirzi parameter
$\gamma\to 0$, while keeping the product $G\gamma$
fixed. This theory has for a long time been conjectured to be
topological.  We prove this statement both in the covariant phase
space formulation as well as in the standard Dirac formulation.  In
the time gauge, the unconstrained phase space of theory admits an
$SU(2)$ connection formulation which makes it isomorphic to the
unconstrained phase space of gravity in terms of Ashtekar-Barbero variables.
Among possible physical applications, we argue that the quantization
of this topological theory might shed new light on the nature of the
degrees of freedom that are responsible for black entropy in loop quantum
gravity.
 
\end{abstract}

\maketitle

\section{Introduction}

The remarkable fact that general relativity can be described in terms
of fields of the kind used in Yang-Mills theories \cite{AshCom}
renewed hope on the possibility of defining a background independent
approach to the canonical quantization of gravity. It was later
realized \cite{BARBERO} that a simple canonical transformation could
be used to replace the (complex) self-dual variables (or Ashtekar
variables) by real $SU(2)$ variables (the so-called Ashtekar-Barbero
variables) more suitable for the definition of the quantization
program. Holst's action was first introduced in \cite{HOLST} as a
covariant formulation of gravity directly leading to the real $SU(2)$
connection formulation upon canonical analysis. The action takes the
following form 
\begin{equation}\label{top1}
	I_{H}=\frac{1}{8\pi G}\int{\hgst (e^I\wedge e^J)\wedge
	F_{IJ}(\omega)}+\frac{1}{8\pi G\gamma}\int{e^I\wedge e^J\wedge
	F_{IJ}(\omega)},
\end{equation}
where $e^I$ is the tetrad 1-forms describing the gravitational field,
$F^{IJ}$ are the curvature 2-forms of a Lorentz connection
${\omega^{IJ}_{\mu}}$, $G$ is Newton's constant, and $\gamma$ is the
co-called Immirzi parameter \cite{IMM}. The $*$ denotes the duality operator
acting on the internal indices $IJKL$.  The first term is the standard
Palatini action of general relativity, while second term can be shown
not to affect the classical equations of motion. The reason for this
is that $\delta_{\omega} I_H=0$ is independent of $\gamma$, and implies
the connection to be the uniquely defined torsion free connection
compatible with $e$: $\omega=\omega(e)$. The second term contribution
to the equation $\delta_e I_H=0$ vanishes identically when evaluated on
$\omega(e)$ due to the Riemann tensor identity $R_{[\mu\nu\rho]\sigma}=0$.
The canonical formulation of the Holst action leads in fact to a
family of $SU(2)$ connection formulations of the phase
space of general relativity labelled by $\gamma$: all of them related by canonical
transformations.  

However, in the quantum theory \cite{lqg} the canonical
transformations relating different connection formulations appear
not to be unitarily implemented. For instance the spectra of 
geometric operators depend on the combination $G\gamma$. Formally speaking, the off shell
contributions of the second term in the Holsts action have a
non trivial effect on amplitudes in the path integral formulation of
quantum gravity. This has an important effect in the computation of black hole 
entropy in LQG \cite{bhe}. There is complete agreement on the universal dependence of entropy 
on fundamental couplings; more precisely, the leading order in the entropy formula
is given by
\be
S_{BH}=\frac{\gamma_0 a_{\va BH}}{4 G\gamma \hbar},
\ee 
where $a_{\va BH}$ is the macroscopic black hole area and $\gamma_0$ is a
dimensionless constant.

This motivates to consider the limit $G\to \infty$ and $\gamma\to 0$
while keeping the product $G\gamma=G_0\gamma_0=constant$. In such a
limit we have \be I_{H}\to I_{0}=\frac{1}{G_0\gamma_0} \int{e^I\wedge
e^J\wedge F_{IJ}(\omega)}, \ee where $I_0$ is a theory thought to be
topological and hence to lack of local physical degrees of freedom. In this
work we will study in detail the classical properties of $I_0$ by
performing its canonical analysis (the treatment of \cite{Alej}
excluded this singular case).  We shall show that the previous limit
is indeed a singular limit where (in the absence of boundaries)
physical degrees of freedom are lost in the limiting procedure.  In
the absence of boundaries $I_0$ is a topological theory.  However, we
will show that non trivial degrees of freedom can arise in the
presence of space time boundaries. Therefore, this singular limit
should be relevant at least for a different understanding of nature of
black hole entropy in LQG.  This is expected to be so from the fact
that the black hole entropy depends on the special combination of
couplings $G\gamma=G_0\gamma_0$, and from the fact that all the
degrees of freedom counted in the calculation of black hole entropy in
LQG are boundary degrees of freedom living on the black hole horizon.

\section{The model}

From now on we concentrate on the study of the model defined by the action $I_0$ which, taking $G_0\gamma_0=1$ and 
putting all the indices, takes the form (this action has been already considered in \cite{merced})
\ba \label{top3}
	&&I_0=\int{e^I\wedge e^J\wedge F_{IJ}(\omega)}.
\end{eqnarray}
The equations of motion of the previous theory are quite simple:
variations $\delta_{\omega}I_0=0$ yield \ba\label{topeom1}
d^{\omega}_{[\mu} (e_{\nu}^Ie_{\rho]}^J)=0 \ \Leftrightarrow\
d^{\omega}(e^I\wg e^J)=0\ \Leftrightarrow\ d^{\omega}e^I=0, \ea
identical to the connection variations of the Hilbert-Palatini
action. Variations of $I_0$ with respect to the tetrad
$\delta_{e}I_0=0$ yield \ba \label{topeom2}
\epsilon^{\alpha\beta\gamma\delta}e_{\beta I} F^{IJ}_{\gamma\delta}=0\
\Leftrightarrow\ e^J\wg F_{IJ}=0. \ea These last field equations are
trivially satisfied once (\ref{topeom1}) hold as a consequence of the
sixteen Riemann tensor identities $R_{[\mu\nu\lambda]\rho}=0$.  This
seems to imply that our theory admits a much larger set of classical
solutions that $I_H$. However, this naive conclusion is indeed
false. The reason is that the action $I_0$ has also a larger group of
local (gauge) symmetries. This can be made clear by a systematic study
of the phase space of the model. In the rest of this section we perform
the canonical analysis of this action. In the next subsection we study
its phase space structure from the covariant phase space perspective
\cite{cov}. In Subsection \ref{HHH} we perform Dirac canonical analysis in a manifestly Lorentz invariant manner.
Finally, in Subsection \ref{TMGG} we study the Dirac formulation in the time-gauge
which allow us to introduce a phase space parametrization in terms of $SU(2)$ connection variables.

\subsection{Analysis in the covariant phase space}\label{CPSS}
Before carrying out formally the canonical analysis, it is worthwhile
to perform a covariant phase space analysis.  By doing this we will
obtain the symplectic potential and symplectic 2-form of our field
theory Eq.(\ref{top3}), also we expect acquiring some rapid
qualitative properties which will provide guidelines for the Dirac canonical
analysis that follows. The following analysis adopts the notations
and conventions in \cite{cov} and the general theory is found in the
references therein.

Let $(\bar{\delta} e^I, \bar{\delta}\omega^{IJ})$ be any variation of
the configuration variables, then the corresponding variation of the
action is \ba &&\bar{\delta}I_0=\int{\left[2e^I\wg F_{IJ}\wg
\bar{\delta}e^J+e^I\wg e^J\wg d^{\omega}\bar{\delta}\omega_{IJ}\right]} 
	\label{cov1} \\
	&&\ \ =\int{\left[2(e^I\wg F_{IJ})\wg
\bar{\delta}e^J-d^{\omega}(e^I\wg
e^J)\wg\bar{\delta}\omega_{IJ}\right]}+\int{d(e^I\wg e^J\wg
\bar{\delta}\omega_{IJ})}, \ea where in the second line, the two terms
in the first integral yield the equations of motion
Eqs.(\ref{topeom1}) and (\ref{topeom2}), while the second integral
gives the symplectic potential \ba\label{cov2}
\Theta(\bar{\delta})=\int_{\Sigma}e^I\wg e^J\wg
\bar{\delta}\omega_{IJ}.  \ea The integration above is carried out on
any time like surface $\Sigma$, and the pull back on $\Sigma$ of
$e^I\wg e^J\wg \bar{\delta}\omega_{IJ}$ is understood. Now let
$\Gamma_{\textrm{cov}}$ be the covariant phase space consisting of all
the solutions of equations of motion (\ref{topeom1}) and
(\ref{topeom2}). Let also $\delta$ be any tangent vector to
$\Gamma_{\textrm{cov}}$, that is, let $(\delta e^I,
\delta\omega^{IJ})$ be any displacement between two neighboring
solutions in $\Gamma_{\textrm{cov}}$. We can calculate the pull-back
of symplectic potential on it, $\Theta(\delta)=\int_{\Sigma}e^I\wg
e^J\wg \delta\omega_{IJ}$. In fact, variating the equations of motion
(\ref{topeom1}) in $\Gamma_{\textrm{cov}}$, we have \ba &&d(\delta
e^I)+\delta \omega^I_{\ J}\wg e^J+\omega^I_{\ J}\wg \delta e^J=0\nnR
\label{cov3} &&\ \ \Rightarrow e^J \wg \delta \omega_{IJ}=d(\delta
e_I)+\omega_{IJ}\wg \delta e^J.  \ea Substituting this into the
pulled-back symplectic potential $\Theta(\delta)$, we have \ba
&&\Theta(\delta)=\int_{\Sigma} e^I\wg e^J\wg (\delta\omega_{IJ})\nnR
&&\ \ \ =\int_{\Sigma}\left[e_I\wg d(\delta e^I)+e^I\wg\omega_{IJ}\wg\delta
e^J\right]
=\int_{\Sigma}\left[e_I\wg d(\delta e^I)+\omega_{JI}\wg e^I\wg\delta e^J\right]
\nnR &&\ \ \ =\int_{\Sigma}\left[e_I\wg d(\delta e^I)-de_I\wg\delta
e^I\right]=-\int_{\Sigma}d(e_I\wg \delta e^I) =-\int_{\partial\Sigma}e_I\wg
\delta e^I, \label{cov4} \ea where the equations of motion
(\ref{topeom1}) is used in the beginning of the third line, and in the
final step $e_I\wg \delta e^I$ is in fact pulled back on $\partial
\Sigma$. Thus the symplectic potential, pulled back on
$\Gamma_{\textrm{cov}}$, turns out to be a total derivative and hence
is a boundary term. The symplectic form $\Omega$, defined as the pull
back to $\Gamma_{\textrm{cov}}$ of the curl of the symplectic
potential, is therefore \ba
\Omega(\delta_1,\delta_2)=-2\int_{\partial\Sigma}\delta_{[1}e_I\wg
\delta_{2]} e^I, \label{cov5} \ea where the infinitesimal
displacements $\delta_1$ and $\delta_2$ in $\Gamma_{\textrm{cov}}$ are
considered also as the tangent vectors of $\Gamma_{\textrm{cov}}$.
We see that in the case $\partial \Sigma=\emptyset$ (or more generally for
restrictive boundary conditions fixing $\delta e=0$) the presymplectic
form (\ref{cov5}) is identically zero.  This implies that all
variations $\delta_1$ or $\delta_2$ are degenerate directions of the
presymplectic form and hence should be regarded as pure gauge.
Thus (locally) all solutions in $\Gamma_{\textrm{cov}}$
are physically equivalent and we have no local degrees of freedom.
This analysis shows that (\ref{top3}) is a topological field theory.

In the cases where $\Sigma$ has boundaries (and depending on the
boundary conditions) the symplectic form can be non-zero. For example
at the presence of a black hole. In such cases, Eq.(\ref{cov5}) has
non-trivial contribution on the horizon.

\subsection{Canonical analysis without time gauge}\label{HHH}

In this section we perform the canonical analysis following Dirac's
method \cite{dirac}.  From now on we assume the spacetime manifold to be of
topology $M=\Sigma\times \R$ where $\Sigma$ is a compact three
manifold. We choose coordinates  $(t,x^a)$ such that the surfaces
$\Sigma_t$ defined by $t=constant$ defines a foliation of $M$, and
$x^a$ with $a=1,2,3$ are local coordinates on $\Sigma_t$ from now on
denoted simply by $\Sigma$. The results presented here were partially
investigated in \cite{maga}. The complete analysis including much more
details than in this paper can be found in \cite{lihui}.

\subsubsection{Primary and secondary constraints}\label{HHH1}

Applying the 3+1 decomposition $e_t^I=Nn^I+N^ae^I_a$ to the action gives
\begin{eqnarray}
	&&I_0=-\frac{1}{2}\int{\epsilon^{abc}\epsilon_{IJKL}(e^I_te^J_a
	\hgst F_{bc}^{KL}+e^I_ae^J_b\hgst F_{tc}^{KL})}\nonumber\\
	&&=\int{\tilde{N}\Pi^b_{IK}\Pi^{cK}_{\ \ \ J}\hgst
	F^{IJ}_{bc}-N^b\Pi^a_{IJ}
	F^{IJ}_{ab}+\omega_t^{IJ}D_a(\Pi^a_{IJ})-\dot{\omega}_a^{IJ}\Pi^a_{IJ}},\label{top4}
\end{eqnarray}
where $\tilde{N}:=-N^2/e$, $e=\textrm{det}(e_{\mu
	I})$ and $\Pi^a_{IJ}=\epsilon^{abc}e_{bI}e_{cJ}$.
By performing the Legendre transformation, one obtains the Hamiltonian 
\begin{eqnarray}
	&&H=\int{\tilde{N}\Pi^a_{IK}\Pi^{bK}_{\ \ \ J}\hgst
	F^{IJ}_{ab}+N^a\Pi^b_{IJ}F^{IJ}_{ab}-\omega^{IJ}_tD_a\Pi^a_{IJ}}+\lambda_a^I
	M^a_I +\lambda^{IJ}_a
	(\Pi^a_{IJ}-\epsilon^{abc}e_{bI}e_{cJ}), \label{top500}
\end{eqnarray} where
	$\tilde N, N^a, \omega_t^{IJ}, \lambda_a^I$, and
	$\lambda_a^{IJ}$ are Lagrange multipliers imposing the primary
	constraints
\begin{eqnarray}
&& M_I^a\approx 0, \label{top6}\\
&& C^a_{IJ}:=\Pi^a_{IJ}-\epsilon^{abc}e_{bI}e_{cJ}\approx 0, \label{top7}\\
&& \cs:=\Pi^a_{IK}\Pi^{bK}_{\ \ \ J}\hgst F^{IJ}_{ab}\approx 0\ \ \textrm{(scalar constraint)}, \label{top8}\\
&& \cv_a:=\Pi^b_{IJ}F_{ab}^{IJ}\approx0\ \ \textrm{(vector constraint)},\label{top9}\\
&& \cg_{IJ}:=D_a\Pi_{IJ}^a\approx0\ \ \textrm{(Lorentz-Gauss law)},\label{top10}
\end{eqnarray}
Here our phase space is parametrized by the canonical pairs $(M^a_I,e_a^I)$,
and $(\Pi^a_{IJ},\omega^{IJ}_a)$.

Now we start studying the consistency conditions of the primary
constraints. The consistency conditions for Eqs.(\ref{top6})
$\dot{M}^a_I\approx 0$ imply
\be\lambda^{JK}_b\epsilon^{abc}e_{cJ}\approx0 \label{top11}, \ee which
can be shown to fix 12 out of the 18 Lagrange multipliers
$\lambda_a^{IJ}$.  This suggests that one can re-combine the 18
constraints $\cp^a_{IJ}$ into two groups: one consists of 6
constraints that commute with $M^a_I$, leading to give 6 secondary
constraints, and the other consists of 12 constraints that do not
commute with $M^a_I$, fixing the 12 multipliers
$\lambda^I_a$. The first
group is given precisely by the (often called) simplicity constraints
\begin{eqnarray}\label{top12}
	&&\Phi^{ab}:=\frac{1}{2}\epsilon^{IJKL}\Pi^a_{IJ}\Pi^b_{KL}=\textrm{Tr}\left(\hgst \Pi^a\Pi^b\right)\approx0,
\end{eqnarray}
The second group is denoted by $\Xi^l\approx0$, $l$ running from 1 to
12. We will calculate the consistency conditions of $\Phi^{ab}$ and
$\Xi^l$ instead of those of $\cp^a_{IJ}$.

To evolve $\Phi^{ab}$ in time, we notice two things that can simplify
the calculation. First, the Gauss-Lorentz law constraints
$\cg_{IJ}\approx 0$ are generators of Lorentz transformation on the
internal indices, so that they commute with any constraints carrying
no internal indices, such as $\Phi^{ab}$.  Second, the vector
constraints $\cv_a\approx0$ and the Gauss law can be combined to give
generators of spatial diffeomorphism
$\tilde{\cv}_a:=\cv_a-\omega^{IJ}_a\cg_{IJ}$, who commute weakly with
$\Phi^{ab}$. The consistency conditions of $\Phi^{ab}\approx0$
can be written in terms of smeared quantities \begin{eqnarray}
&&\dot \Phi^{ab}[\lambda_{ab}]\approx \{\Phi^{ab}[\lambda_{ab}],\cs[\tilde{N}]\}=\int\int\epsilon^{IJKL}\lambda_{ab}\Pi^b_{KL}\{\Pi^a_{IJ},
F^{MN}_{cd}\}\tilde{N}\left(\hgst \Pi^c\Pi^d\right)_{MN}\nonumber\\
&&\ \ =4\int{\lambda_{ab}\hgst \Pi^{bIJ}D_c\left(\tilde{N}\left(\hgst
\Pi^c\Pi^a\right)_{IJ}\right)}
\approx-4\int{\tilde{N}\lambda_{ab}\textrm{Tr}\left(\hgst \Pi^a\hgst
\Pi^cD_c\Pi^b\right)}\nonumber\\ \label{top13} &&\ \
=4\int{\tilde{N}\lambda_{ab}\textrm{Tr}\left(\Pi^a\Pi^cD_c\Pi^b\right)}:=\chi^{ab}[\tilde{N}\lambda_{ab}],
\end{eqnarray}
where $\Phi^{ab}[\lambda_{ab}]=\int \lambda_{ab} \Phi^{ab}$ and
similarly for $\cs[\tilde N]$ and $\chi^{ab}[\tilde{N}\lambda_{ab}]$.
Here we used the Gauss law constraint and the fact that
$\textrm{Tr}(\Pi^{(a}\Pi^{|c|}\Pi^{b)})\approx0$ by virtue of
Eq.(\ref{top7}). This leads to 6 secondary constraints
\begin{equation}\label{top14}
	\chi^{ab}:=\textrm{Tr}\left(\Pi^{(a}\Pi^{|c|}D_c\Pi^{b)}\right)
	=-\textrm{Tr}\left(\hgst \Pi^{(a}\hgst \Pi^{|c|}D_c\Pi^{b)}\right)\approx0.
\end{equation}
We do not bother to care about the exact expression of $\Xi^l$. The
consistency conditions $\dot \Xi^l$ fix the multipliers of constraints
Eq.(\ref{top6}), the 12 $\lambda^I_a$ and no secondary constraint
arises.  As for the 12 multipliers of $\Xi^l$, they are in fact just
those that are fixed in Eq.(\ref{top11}). Thus $\Xi^l\approx0$ fall in
the second class together with $M^a_I\approx0$, and they discard the
12 degrees of freedom carried by $e^I_a$. The evolution of $\chi^{ab}$ 
does not lead further constraints.

\subsubsection{Reducibility of the constraints}\label{HHH2}

At this stage, a naive counting would yield a negative number of
degrees of freedom.  This is a clear indication that not all
constraints are independent: there is reducibility in the constraint
system.  In fact we will now prove that the scalar and vector
constraints are in fact implied by the Gauss-Lorentz law and the
secondary constraints Eq.(\ref{top14}).
To see this let us express the relevant constraints ($\cg_{IJ}$,
$\chi^{ab}$, $\cs$ and $\cv_a$) in terms of the tetrad components,
with the help of Eq.(\ref{top7}). In particular, on one hand for
$\cg_{IJ}$ and $\chi^{ab}$,
\begin{eqnarray}
	&&\cg_{IJ}\approx
	D_a\left(\epsilon^{abc}e_{bI}e_{cJ}\right)=\epsilon^{abc}e_{b[I}D_ae_{cJ]},
	\label{toop1} \\ &&\chi^{ab}=\hgst \Pi^{(aK}_I\hgst
	\Pi^{|c|}_{KJ}D_c\Pi^{bJI}\approx
	e^2\left(e^t_Ie^{(aK}-e^{tK}e^{(a}_I\right)\left(e^t_Ke^{|c|}_J-e^t_Je^{|c|}_K\right)
	D_c\left(\epsilon^{b)fg}e^J_fe^I_g\right)\nonumber\\ &&\ \ \ \
	=e^2\left(e^t_Ie^c_Jg^{t(a}-e^t_Ie^t_Jg^{c(a}-g^{tt}e^c_Je^{(a}_I+g^{tc}e^t_Je^{(a}_I\right)
	D_c\left(\epsilon^{b)fg}e^J_fe^I_g\right)\nonumber\\ &&\ \ \ \
	=e^2\left(e^t_Ig^{t(a}-g^{tt}e^{(a}_I\right)\epsilon^{b)cd}D_ce_d^I=e^2N^{-2}
	\left(e^t_IN^{(a}+e^{(a}_I\right)\epsilon^{b)cd}D_ce_d^I\
	. \label{toop2}
\end{eqnarray}
Here $g^{\mu\nu}:=e^{\mu I}e^{\nu}_I$ is the inverse spacetime metric,
and it is related with the lapse and the shift by $g^{at}=N^a/N^2$ and
$g^{tt}=-1/N^2$ (see \S2.3 of \cite{Peldan}).  One can show\footnote{A
key step in showing that the transformation matrix from (\ref{toop1}) and (\ref{toop2})
is non degenerate is to write down the inverse tetrad component in terms of the
tetrad component:
$e^a_I=\frac{1}{2e}\epsilon^{abc}\epsilon_{JIKL}e^J_te^K_be^L_c.$}
that the previous twelve constraints are equivalent to
\begin{equation}\label{top16}
	\cc^{aI}:=\epsilon^{abc}D_be_c^I\approx0.
\end{equation}
Applying $D_a$ to these constraints, we obtain
\begin{equation}\label{top17}
	D_a\cc^{aI}=\epsilon^{abc}D_aD_be^I_c=\epsilon^{abc}e_{cJ}F^{IJ}_{ab}\approx0.
\end{equation} On the other hand, the constraints $\cs$ and $\cv_a$,
can be written as
\begin{eqnarray}\label{toop5}
	&&\cs=\hgst \Pi^a_{IK}\Pi^{bK}_{\ \ J}F^{IJ}_{ab}\approx
	e\left(e^t_Ie^a_K-e^t_Ke^a_I\right)\epsilon^{bcd}e^K_ce_{dJ}F^{IJ}_{ab}=-e\epsilon^{abc}e^t_Ie_{cJ}F^{IJ}_{ab},
	\\
	\label{toop6}&&\cv_a=\Pi^b_{IJ}F^{IJ}_{ab}\approx\epsilon^{bcd}e_{cI}e_{dJ}F^{IJ}_{ab}=\frac{1}{2} \epsilon^{bcd}e_{cI}e_{dJ}\epsilon_{abf}\epsilon^{fgh}F^{IJ}_{gh}=-e_{aI}\epsilon^{bcd}e_{bJ}F^{IJ}_{cd},
\end{eqnarray}
both of which vanish as a consequence of (\ref{top17}).  Therefore,
the constraints $\cs$ and $\cv_a$ are implied by the constraints
(\ref{top7}), (\ref{top10}), and (\ref{top14}). Thus they can be safely removed
from the Hamiltonian (\ref{top500}). This operation preserves the
constraint surface as well as the trajectories of motion, at the
harmless cost of certain modifications of multipliers of $\cg_{IJ}$
and $\chi^{ab}$. However certainly one has to add $\chi^{ab}$ to the
Hamiltonian, which now reads
\begin{eqnarray}
	&&H=\int\kappa^{IJ}\cg_{IJ}+\kappa_{ab} {\chi}^{ab}
	+\gamma_{ab}\Phi^{ab}+\gamma_l\Xi^l+\lambda^I_aM^a_I.\label{top18}
\end{eqnarray}

Now we are ready to classify the constraints. Our analysis so far
shows that $\cg_{IJ}, \Phi^{ab}$ and ${\chi}^{ab}$ are first class;
while $M^a_I$ and $\Xi^l$ are second class.  Thus for the 60
dimensional unconstrained phase space parametrized by $(M^a_I,e_a^I)$,
and $(\Pi^a_{IJ},\omega^{IJ}_a)$ we have 18 first class constraints
and 24 second class constraints which yields zero local degrees of
freedom as expected from the analysis of Subsection \ref{CPSS}.

Further insight into the nature of this topological model will be gained by repeating this analysis using the
partial gauge fixing of the Lorentz symmetry known as the time gauge. This will reduce
the internal gauge group from $SO(3,1)$ to $SO(3)$, and will make the relationship with gravity
more explicit.

\subsection{Canonical analysis under time gauge}\label{TMGG}

\subsubsection{The Hamiltonian and the primary constraints under time gauge}\label{TMGG1}
To redo the analysis under time gauge, let us return to the
Hamiltonian (\ref{top500}). The time gauge condition is defined by
identifying the zero-th component of the tetrad $e_{\mu 0}$, with
$n_{\mu}=(-N,0,0,0)$, the the co-normal of the space-like
hyper-surfaces of $3+1$ foliation of spacetime\footnote{Here we need to
choose $e^0_{\mu}=n_{\mu}=(-N,0,0,0)$ instead of letting $e_{\mu
0}=n_{\mu}$ because of the convention $\textrm{det}(e_{\mu I})>0$,
which is chosen to let $e^0_{\mu}$ to be future pointing.}. This is
equivalent of imposing $n_Ie^I_a=0$, where $n_I=e_{\mu I}n^{\mu}$. One
can also prove that under this condition, $n_I=(1,0,0,0)$. Therefore,
we can impose the time gauge condition by adding to the list of
primary constraints Eqs.(\ref{top6})---(\ref{top10})
\begin{equation}\label{TG1}
	e_a^0\approx0.
\end{equation}
They give 6 second class constraints together with $M^a_0\approx0$
which can be solved directly in order to get rid of $e_a^0$ and
$M^a_0\approx0$ from the analysis. In this process the phase space is
reduced to  the canonical pairs $(M_i^a, e^i_a)$ and $(\Pi_{IJ}^a,
\omega^{IJ}_{a})$, and the action (\ref{top4}) becomes:
\begin{eqnarray}
	I_0
=\int{-\frac{N}{2}\frac{ \epsilon_{i}^{\ jk} E_j^aE_k^b F^{i0}_{ab}}{ \sqrt{{\rm det}E}}-N^b E^a_i
	F^{i}_{ab}+\omega_t^{i0} D_a(\Pi^a_{i0})+\omega_t^{ij} D_a(\Pi^a_{ij})
-\dot{A}_a^{i}E^a_{i}},
\end{eqnarray}
where we used the definitions $E^a_i:=\frac{1}{2}\epsilon^{\
jk}_i\Pi^a_{jk}$, and $A^i_a:=-\frac{1}{2}\epsilon^i_{\
jk}\omega_a^{jk}$. If in addition we define 
$K^i_a:=\omega^{0i}_a$ the previous expression becomes
\begin{eqnarray}
	I_0
=\int{-E^a_{i} \dot{A}_a^{i}}+\Pi_i^a\dot K^i_a- H,\label{topytop}
\end{eqnarray}
where the Hamiltonian takes the (perhaps) more familiar form \be
H=\int \frac{N}{2}\frac{ \epsilon_{i}^{\ jk} E_j^aE_k^b
{\sD}_aK^i_{b}}{ \sqrt{{\rm det}E}}+N^b E^a_i \sF^{i}_{ab}+N^i
\epsilon_{ijk}E^{aj}K^k_a +M^i {\mfs D}_aE^a_i+\lambda_a^i C^a_i
+\rho_a^i M^a_i +\gamma_a^i\Pi_{i0}^a, \ee where $\sD$ and $\sF$ are
the covariant derivative and curvature of the $SU(2)$ connection
$A_a^i$, and $N,N^a,N^i,M^i,\lambda^a_i,\rho^a_i$, and $\gamma_a^i$
are Lagrange multipliers.  The Poisson brackets among the basic variables
are
\begin{equation}\label{TG9}
	\{E^a_i(x),A^j_b(y)\}=\delta^a_b\delta^j_i\delta^3(x,y),\ \ \{K^i_a(x),\Pi^b_{0j}(y)\}=\delta^b_a\delta^i_j\delta^3(x,y),
\end{equation}
and the primary constraints are
\ba
&& M^a_i \approx 0\\
&& C^a_i:=E^a_i-\frac{1}{2} \epsilon^{abc}\epsilon_{ijk}e_b^je_c^k\approx 0 \label{top7p}\\
&& \cs:=\frac{ \epsilon_{i}^{\ jk} E_j^aE_k^b {\sD}_aK^i_{b}}{ \sqrt{{\rm det}E}} \approx 0\ \ \textrm{(scalar constraint)}, \label{top8p}\\
&& \cv_a:=E^a_i \sF^{i}_{ab}\approx0\ \ \textrm{(vector constraint)},\label{top9p}\\
&& \cg_{i}:={\mfs D}_aE^a_i \approx0\ \ \textrm{($SO(3)$ Gauss law)},\label{top10p}\\
&& B_{i}:= \epsilon_{ijk}E^{aj}K^k_a \approx 0\ \ ,\label{top10pp}\\
&& \Pi^a_{i0}\approx 0.
\ea

\subsubsection{Secondary constraints under time gauge}\label{TMGG2}

The consistency condition $\dot M^a_i\approx 0$ implies
\begin{eqnarray}
\lambda^j_b\epsilon^{abc}\epsilon_{ijk}e_c^k+\cdots \approx0
\ea
which fixes the nine Lagrange multipliers $\lambda^i_a$. The consistency condition $\dot C^a_i\approx 0$ gives
\ba -\rho^j_b\epsilon^{abc}\epsilon_{ijk}e_c^k+ \cdots\approx0,\label{TG24}
\end{eqnarray}
which fixes the Lagrange multipliers $\rho_a^i$.  The consistency
conditions $\dot \Pi^a_{i0}\approx 0$ are best understood if we 
split the nine components of $\Pi^a_{i0}$ as follows \cite{Alej} 
\begin{eqnarray}\label{TG42}
	&& \Pi^i:=\epsilon^{ijk}e_{aj}\Pi^a_{0k},\ \ \Pi_{ij}:=e_{a(i}\Pi^a_{0j)}=\Pi_{ji},
\end{eqnarray}
Now $\dot \Pi_{ij}\approx 0$ implies six secondary constraints 
\be
S_{ij}:= \epsilon^{abc}e_{a(i}{\mfs D}_be_{cj)}\approx 0,  \label{TG43}
\ee while $\dot \Pi_i$ implies
\begin{eqnarray}
N^i-{{}^3\!e^{-1}E^{ai}\partial_aN'}+\cdots \approx0, \label{TG44}
\end{eqnarray}
which fixes the three Lagrange multipliers $N^i$.  At this stage an
important remark is in order. Notice that the six constraints
$S_{ij}=0$ together with the three Gauss law three $\sD_aE^a_i\approx
0$ (\ref{top10p}) are equivalent to the nine
$\epsilon^{abc}\sD_{b}e_{c}^i\approx 0$ which in turn can be more conveniently written as
\begin{equation}\label{TG45}
\mathbb{D}^i_a:=A^i_a-\Gamma^i_a(E)\approx0,
\end{equation}
where $\Gamma^i_a$ is the spin connection compatible with the triad $e_a^i$.
Therefore, the secondary constraints (\ref{TG43}) and the (\ref{top10p}) can be replaced by (\ref{TG45}).

\subsubsection{Reducibility of the constraints}\label{TMGG3}
Same as in the direct analysis, \S\ref{HHH2}, we can prove that the
scalar constraint and the vector constraints are implied by other constraints and hence redundant.

Due to (\ref{TG45})
${\mfs D}_{[a}e_{b]}^i\approx 0$ the scalar constraint can be re-written as 
\begin{eqnarray}\label{TG50}
	\cs=\frac{ \epsilon_{i}^{\ jk} E_j^aE_k^b {\sD}_aK^i_{b}}{ \sqrt{{\rm det}E}} \approx 
{\sD}_a (\frac{ \epsilon_{i}^{\ jk} E_j^aE_k^b K^i_{b}}{ \sqrt{{\rm det}E}})\approx 0
\end{eqnarray}
where in the last equality we have used (\ref{top10pp}). The previous equation tell us that the scalar
constraint is in fact  implied by the constraints (\ref{TG45}) and
(\ref{top10pp}), or equivalently by (\ref{top10pp}), (\ref{TG43}) and
(\ref{top10p}).  A similar thing happens for the vector constraint. We
first observe that ${\mfs D}_{[a}e_{b]}^i\approx 0$ implies
$\epsilon^{abc}{\mfs D}_a {\mfs D}_{b}e_{c}^i\approx 0$ from which we
obtain (using the definition of the curvature strength) $\epsilon^{abc}\epsilon^i_{\ jk}{\sF}^j_{ab}e^k_c\approx0$. Using 
the constraint (\ref{top7p}) one shows in a line that this implies
\begin{eqnarray}\label{TG53}
{\sF}^j_{ab} E^a_iE^b_j={\cal V}_aE^a_i\approx0.
\end{eqnarray}
Using the (assumed) invertibility of $E^a_i$ we conclude that the
vector constraints ${\cal V}_b\approx 0$ are implied by the constraints
(\ref{TG45}) and (\ref{top7p}).  There are no more redundant
constraints.  Eliminating the redundant constraints the Hamiltonian
can be written as \be H_T=\int [N^i \epsilon_{ijk}E^{aj}K^k_a+M^i {\mfs
D}_aE^a_i+\alpha^{ij} S_{ij}+\lambda^a_i
(e_a^i-\frac{\epsilon_{abc}E_j^bE_k^b\epsilon^{ijk}}{2 \sqrt{{\rm
det}(E)}}) +\rho_a^i M^a_i +\gamma^i\Pi_{i}+\gamma^{ij}
\Pi_{ij}],\label{h1} \ee where we have added the secondary constraint
$S_{ij}$ with its Lagrange multiplier $\alpha^{ij}$ to the total
Hamiltonian, and $\Pi_i$ and $\Pi_{ij}$ were defined in (\ref{TG42}).
Equivalently we can write \be H_T=\int [N^i \epsilon_{ijk}E^{aj}K^k_a+
\alpha^a_i(A_a^i-\Gamma_a^i(E))+\lambda^a_i
(e_a^i-\frac{\epsilon_{abc}E_j^bE_k^b\epsilon^{ijk}}{2 \sqrt{{\rm
det}(E)}}) +\rho_a^i M^a_i +\gamma^i\Pi_{i}+\gamma^{ij}
\Pi_{ij}],
\label{h2} \ee where we have replaced the Gauss
law and $S_{ij}$ by the equivalent condition (\ref{TG45}).

\subsubsection{Classification of constraints and solution of second class constraints}\label{TMGG4}
There are no further secondary constraints, we can thus proceed to
their classification.  Recalling the notation
$B_i:=\epsilon_{ijk}E^{aj}K^k_a$, and
$C_a^i=e_a^i-{\epsilon_{abc}E_j^bE_k^b\epsilon^{ijk}}/{(2 \sqrt{{\rm
det}(E)})}$ (notice that instead of $C^a_i$ defined in (\ref{top7p})
we are using its inverse for convenience).  Their algebra is
summarized in the following matrix
\begin{eqnarray}
	&&\hspace{1.55cm}M^a_i \hspace{1cm} C_a^i
	\hspace{1.1cm} B_i \hspace{1.2cm} \Pi^i
	\hspace{0.42cm} \hspace{0.55cm} \mathbb{D}^a_i \hspace{0.68cm}
	\Pi_{ij} \nonumber \\ &&\begin{array}{c} M^b_j \\
	C^j_b \\ B_j \\ \Pi^j \\ \mathbb{D}^b_j
	\\ \Pi_{kl}
	\end{array} 
	\left(\begin{array}{cccc|cc}
		0 & -\delta^i_j\delta^b_a\delta^3_{xy} & 0 & 0 & 0 & 0 \\
		\delta^j_i\delta^a_b\delta^3_{xy} & 0 & 0 & 0 & \{C^j_b,\mathbb{D}^a_i\} & 0 \\ 
		0 & 0 & 0 & -2 e\delta^i_j\delta^3_{xy} & \{B_j,\mathbb{D}^a_i\} & 0 \\ 
		0 & 0 & 2 e\delta^j_i\delta^3_{xy} & 0 & 0 & 0 \\ \hline
		0 & \{\mathbb{D}^b_j,C_a^i\} & \{\mathbb{D}^b_j,B_i\} & 0 & 0 & 0 \\
		0 & 0 & 0 & 0 & 0 & 0
	\end{array}\right),\label{TG60}
\end{eqnarray}
where values are to be read in the weak sense, and we have used the known fact that (in the absence of boundaries)
$\Gamma^i_a(E)=\delta F/\delta E^a_i$ for $F:=\int{E^a_i\Gamma^i_a}$, which implies
\begin{eqnarray}\label{TG70}
	\{\mathbb{D}^a_i,\mathbb{D}^b_j\}=-\{\Gamma^a_i,A^b_j\}-\{A^a_i,\Gamma^b_j\}=-\frac{\delta
	\Gamma^a_i}{\delta E^j_b}+\frac{\delta \Gamma^b_j}{\delta
	E^i_a}=0.
\end{eqnarray}
The matrix Eq.(\ref{TG60}) implies that the $\Pi_{ij}$ are first class, the $\mathbb{D}^a_i$ can be made into first class
by the addition of a suitable combination of the constraint in the upper left block consisting of 
$M^a_i$, $C_a^i$, $B_i$, and $\Pi^i$ which are second class.
Thus we have 15 first class constraints and 24 second class
constraints. The phase space is spanned by $(e^i_a,A^i_a,K^i_a)$ and
their conjugate momenta so that it has 54 dimensions. Therefore, there are $54/2-15-24/2=0$ physical
degrees of freedom. This result is consistent with the counting
of the previous Subsections (\ref{CPSS}) and (\ref{HHH}).

\subsubsection{A partial reduction}

In order to compare our model with the description of general
relativity in terms of Ashtekar-Barbero variables it will be
convenient to resolve the second class constraints above and gauge-fix
the gauge symmetry generated by the first class constraints
$\Pi_{ij}$.  The first step is immediate as far as the constraints
$C_a^i=0$ and $M^a_i=0$ are concerned. One just substitutes
$e^i_a$ using $C_a^i=0$ everywhere and sets $M^a_i=0$. By doing
so the triad variables and their conjugate momenta are excluded from
the phase space. Similarly for $B_i=0$ and $\Pi^i=0$ which removes
three of the degrees of freedom in $K^i_a$ (namely the $B^i$) and their conjugate
momenta.  In this way we are left with six remaining degrees of freedom in
$K_a^i$.  More precisely, these are given by
$K^{ij}:=E^{a(i}K_a^{j)}$. We can get rid of them by imposing
the gauge fixing condition \be K^{ij}=0\label{55} \ee which fixes the gauge
freedom generated by the six $\Pi_{ij}$. The reduced system is described by the action
\be
I_{red}[A,E]=\int dt\int_{\Sigma} \left[E^a_i\dot A^i_a-N^i\sD_aE^a_i-\alpha^{ij}S_{ij}\right],
\ee
or equivalently
\be\label{60}
I_{red}[A,E]=\int dt\int_{\Sigma} \left[E^a_i\dot A^i_a-\alpha^{a}_i(A_a^i-\Gamma_a^i)\right].
\ee
The constraints are manifestly first class and the previous actions define
a background independent $SU(2)$ connection gauge theory with no local degrees of freedom.

\section{Conclusions}

We have performed the canonical analysis of the theory (\ref{top3}) in
three alternative ways.  First the covariant phase space formulation
of Subsection {\ref{CPSS}} allows us to quickly learn that the theory
is topological in the absence of boundaries. In Subsection \ref{HHH}
we perform the Dirac analysis and obtain all the constraints and
their classification. The counting of degrees of freedom is in
agreement with the results of the covariant phase space
formulation. However, second class constraints turn out to be rather
complicated. The comparison with gravity in the Ashtekar-Barbero
formulation suggested the analysis of the formulation of the field theory in
the time gauge. With this partial gauge fixing, we find a surprisingly
simple expression for the action of the model expressed in terms of a
canonical pair $(A_a^i, E^a_i)$ of an $SU(2)$ connection and its
conjugate non Abelian electric field satisfying the usual Gauss (first
class) constraints $\sD_aE^a_i\approx 0$ plus six additional (first
class) constraints stemming from 4-diffeo invariance of the original
action plus two additional gauge symmetries that---from the perspective of the Holst
action---kill the
would-be-gravity degrees of freedom. These nine (first class) constraints can be concisely expressed 
by the conditions
\[ A_a^i-\Gamma^i_a\approx 0\]
which are manifestly first class. From this fact, one could 
have had guessed at posteriori that action (\ref{60}) is a consistent 
gauge theory with no local degrees of freedom. The extra merit of our analysis is to show that
(\ref{60}) comes indeed from (\ref{top3}).

We would like to stress a novel feature of the theory studied here. On the one hand it is a very simple model
as it does not have any local degrees of freedom in the absence of boundaries. In this respect it shares a place 
with other topological theories in 4d such as BF theory. 
 On the other hand, and this is a unique feature of this model, the field content of the theory is exactly the same 
 as the one of general relativity in the first order formulation: namely the gravitational field $e_a^I$ and the Lorentz connection $\omega^{IJ}_a$.
Moreover, the phase space of the theory can be described by $SU(2)$ connection variables just as in the gravity case.
This may make this theory an interesting playground to test ideas  
relevant for gravity in 4d in a simpler context (in particular when it concerns quantization) .

Notice that all the quantization techniques of loop quantum gravity can be directly imported 
to this simple theory: the definition of the kinematical Hilbert space, the quantization of geometric operators such as 
area and volume, and the quantization techniques of Thiemann for the promotion of the constraints to quantum operators.
For example one could promote the nine constraints above to operators by replacing Poisson brackets by commutators in
the classical identity
\[ A_a^i-\Gamma^i_a=-2\{\{H_{E}(1),V\},A_a^i\},\]
were $H_{E}(1)$ is the so-called Euclidean Hamiltonian (see for instance eq. 10.3.7
and 10.3.16 in Thiemann's book \cite{lqg}).

Our argument given in the introduction suggests that the theory
studied here should play an important role in understanding the origin
of black hole entropy.  In the standard treatment of black hole
entropy in LQG one quantizes gravity in a spacetime with a boundary at
the location of the black hole event horizons (with appropriate
boundary conditions defining a so-called isolated horizon \cite{IH}).
Our analysis implies that in the limit $G\to \infty$ and $\gamma\to 0$ with $G\gamma$ held 
constant discussed in the introduction black hole entropy remains fixed, 
while the gravitational degrees of freedom in the bulk disappear.
The results of section \ref{CPSS} tell us that degrees of freedom might remain at the boundary.
But it is precisely only boundary degrees of freedom that enter the standard calculation of black hole entropy.
Therefore, all this strongly suggests that the origin of black hole entropy can, in this sense,
be associated with excitations of our simple model.

\section{Acknowledgments}

We would like to thank the remarks and questions raised by an anonymous referee which 
have led to the improvement of this work. This work was supported in part by CONACYT, Mexico, Grant Numbers 56159-F
and 79629 (sabbatical term). MM thanks the {\it Centre de Physique
Th\'eorique} at Luminy, Marseille for all support and facilities provided
for the realization of his sabbatical term. AP Thanks the support of the 
{\em Intitut Universitaire de France} and grant ANR-06-BLAN-0050.  In the appendix we discuss this point further and we exhibit a simple example of boundary condition leading to 
local degrees of freedom at the boundary.

\begin{appendix}
\section{Boundary degrees of freedom}

In this appendix we explicitly exhibit examples of how the system described in this paper can have local 
degrees of freedom if the space-time considered contains a boundary where, by defining 
assumption of the variational principle, fields are allowed to vary while appropriate boundary 
conditions are satisfied.  In the first example we simply start from equation (\ref{cov5}) and require
some extra conditions on the one forms $e^I$ on the boundary. A possible way to define natural boundary 
conditions is to start from the symmetry content we want  the theory to have at the boundary. We will assume that
boundary manifold is foliated by a preferred family of two-surfaces $H$ and that the space time foliation is arbitrary 
in the bulk but it is restricted to coincide with the preferred foliation of the boundary at the boundary, namely $H=\partial \Sigma$. 
We will work in the time gauge $e^0=0$  and require $SU(2)$ local transformations of the triad at the boundary---from now on denoted $G(SU(2))$---as well as ${\rm Diff}(H)$ to be gauge symmetries of the
boundary fields. This implies that the pre-symplectic structure (\ref{cov5}) will have to have null vectors associated to these
transformations. The symmetry requirement will define for us boundary conditions for the given field content. Notice also that this is precisely the 
symmetry content of the isolated horizon boundary condition \cite{bhe}.

Let us start with $SU(2)$ transformations. Under an infinitesimal $SU(2)$ transformation parametrized by the field $\alpha \in su(2)$ 
the triad transforms as $\delta_{\alpha} e^i=[\alpha,e]^i$. This transformation is a gauge symmetry if for all $\alpha$ and arbitrary $\delta\in \Gamma_{\rm cov}$ the equation
$\Omega(\delta_{\alpha},\delta)=0,$ namely
\be
-\Omega(\delta_{\alpha},\delta)=\int_{H=\partial \Sigma} \delta_{\alpha}e^i\wedge \delta e_i=\int_{H} [\alpha,e]^i\wedge \delta e_i=\frac{1}{2}\int_{H} \delta(\epsilon_{ijk} \alpha^j e^k\wedge e^i)=0.
\ee
The previous equation tell us that, given the present field content, in order to preserve $SU(2)$ gauge invariance at the boundary we must impose the (zero area) boundary condition
\be
\Sigma_i=\epsilon_{ijk} e^j\wedge e^k=0.\label{a=0}
\ee
This boundary condition is certainly inappropriate for studies in the context of the black hole entropy, we will describe below a different alternative more suitable for
 such context. Notice that only two out of the tree constraints $\Sigma^i=0$ are really independent. The next gauge symmetry we would like to impose is ${\rm Diff}(H)$. 
 Under  an infinitesimal diffeomorphisms parametrized by a vector field $v\in T(H)$ the triad transforms as $\delta_{v}e^i=d(v\inter e^i)+v\inter de^i$. Similarly to  the previous case, the
 requirement $\Omega(\delta_{v},\delta)=0$ for all $\delta\in \Gamma_{\rm cov}$ becomes:
\ba \nonumber &&
\Omega(\delta_{v},\delta)=\int_{H} \delta e^i\wedge \delta_v e_i=\int_{H} \delta e_i\wedge (d(v\inter e^i)+v\inter de^i)=\\ \nonumber && 
=\int_{H}  d(\delta e_i) \wedge (v\inter e^i) -d(\delta e_i \wedge (v\inter e^i)) + {\delta e_i\wedge (v \inter de^i)}= \\ &&=\int_{H} \delta( de_i  (v\inter e^i))=0,
\ea
where we have used the fact that $\delta e_i\wedge (v \inter de^i)=(v \inter \delta e_i )\wedge  de^i$ in the last term of the second line, and have assumed $\partial H=0$ in the last line.
 At first sight one would the conclude that ${\rm Diff}(H)$ are gauge symmetries of the system if and only if the following vector constraint is satisfied
 \be
 V_a=   e_{ai}de_{bc}^i\epsilon^{bc}=0;
 \ee
 however, if we recall the bulk equation of motion (\ref{topeom1}), the time gauge, and the gauge condition (\ref{55}), we see that the previous constraint is implied by $\Sigma^i=0$ as 
 $V_a=e_a^i\Gamma_b^j e_c^k\epsilon_{ijk}\epsilon^{bc}=-\Gamma^j\Sigma_j=0$. Therefore we conclude that the only constraints on boundary fields, necessary to preserve the symmetry content
 required in our example is given by (the two independent components of) the vanishing area constraint (\ref{a=0}). It is immediate to check that the vanishing area constraints are indeed first class. 
 The unconstrained phase space is parametrized by the 6 local fields $e_a^i$ which implies a reduced phase space parametrized by two local fields, i.e. the system defined in this example has
 one local degree of freedom on the boundary. Notice that this is a kind of generalization of the Husain-Kuchar model \cite{HK}, as those studied in \cite{merced}.

We have seen that with the field content given above the symmetry requirement $G(SU(2))\rtimes {\rm Diff}(H)$---the symmetry group of isolated horizons---implies that area vanishing constraint $\Sigma^i=0$ and therefore this system cannot accommodate in any suitable way the black hole system that motivated the study of the theory considered in this work. However, this conclusion can be circumvented if one allows for additional field content at the horizon. In particular, if in addition one allows for an $SU(2)$ connection $A^i$ to be an independent degree of freedom at the boundary then the considerations that lead to the result of \cite{kaen} imply that, if the isolated horizon boundary condition $\Sigma^i=-(a/\pi) F^i(A)$ is satisfied, then $G(SU(2))\rtimes {\rm Diff}(H)$ of the enlarged field system is gauge symmetry group of the system, and the presymplectic structure becomes
\be
\Omega(\delta_{1},\delta_{2})=\int_{H} \frac{a}{2 \pi}\delta_1 A_i\wedge \delta_2 A^i - \delta_1 e^i\wedge \delta_2 e_i,
\ee
where the first term is a boundary term added in order to preserve gauge invariance in the presence of a non vanishing  boundary area while the
second term is the boundary term coming from the bulk. It is possible that the detail study of the quantization of this model could shed light on the nature of BH entropy 
in LQG. However, even when a lot is known about the quantization of the first term (given by an $SU(2)$ Chern-Simons theory of the kind studied in \cite{witten}) 
this is not an easy task as it would require the background independent quantization of the second term defining the dynamics of the $e_a^i$ field about which, to our knowledge, little is known. We hope to be able to deepen the understanding of this model in the future.

\end{appendix}

\end{document}